# Perceptron Collaborative Filtering

Arya Chakraborty

*12th grade student, Delhi Public School Ruby Park, Kolkata, India*
aryastlawrence@gmail.com
aryachakraborty2005@gmail.com

*Abstract*— While multivariate logistic regression classifiers are a great way of implementing collaborative filtering - a method of making automatic predictions about the interests of a user by collecting preferences or taste information from many other users, we can also achieve similar results using neural networks. A recommender system is a subclass of information filtering system that provide suggestions for items that are most pertinent to a particular user. A perceptron or a neural network is a machine learning model designed for fitting complex datasets using backpropagation and gradient descent. When coupled with advanced optimization techniques, the model may prove to be a great substitute for classical logistic classifiers. The optimizations include feature scaling, mean normalization, regularization, hyperparameter tuning and using stochastic/mini-batch gradient descent instead of regular gradient descent. In this use case, we will use the perceptron in the recommender system to fit the parameters i.e., the data from a multitude of users and use it to predict the preference/interest of a particular user.

## I. A Brief History and Introduction

### A. Neural Networks

Neural network in machine learning is a model made up of artificial neurons or nodes. The connections between the nodes are modelled as weights. A positive weight reflects an excitatory connection while a negative weight reflects an inhibitory connection. The inputs are modified by the weights and summed; an activity known as linear combination. An activation function at the end controls the amplitude of the output i.e., brings the output in a desirable range – usually 0 to 1 or -1 to 1.

In 1943, Warren McCulloch and Walter Pitts from the University of Illinois and the University of Chicago published research that analysed how the brain could produce complex patterns and could be simplified down to a binary logic structure with only true/false connections. Frank Rosenblatt from the Cornell Aeronautical Laboratory was credited with the development of the perceptron in 1958. His research introduced weights to McColloch's and Pitt's work, and Rosenblatt leveraged his work to demonstrate how a computer could use neural networks to detect imagines and make inferences.

The next step in the development of neural networks came in 1982 with the development of 'Hopfield Networks' by John Hopfield. A Hopfield network is a fully interconnected recurrent neural network where each unit is connected to every other unit. It behaves in a discrete manner and produces distinct outputs in generally binary (0/1) form or in bipolar (-1/1) form. In a recurrent neural network, the outputs of the neurons are again fed into the network as 'memory' that improves the current output and input of the network.

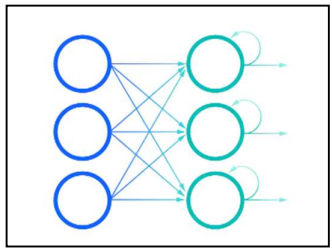

Fig. 1 Diagram of Recurrent Neural Network (Source: https://www.ibm.com/cloud/learn/recurrent-neural-networks)

The backpropagation algorithm, which nowadays forms the basis of the neural networks, though independently discovered many times earlier, the modern form was proposed by Yann LeCun in 1987. Paul Werbos was the first in US to propose that backpropagation could be used for neural nets after deep analyzation in his 1974 dissertation. The Werbos method was rediscovered in 1985 by Parker and in 1986 by David Everett Rumelhart, Geoffrey Everest Hinton and Ronald J. Williams. However, the basics of continuous backpropagation were derived in the context of 'Control Theory' by Henry J. Kelley and Arthur E. Bryson in 1960 and 1961 respectively

After a period of minor developments during the 2000s, the resurrection of neural networks began in 2010s because they benefitted from the cheap, powerful GPU-based computing systems. This is much more noticeable in the fields of speech recognition, machine vision, natural language processing (NLP) and language structure learning. Other use cases for neural networks include facial recognition, stock market prediction, social media, aerospace, health etc.

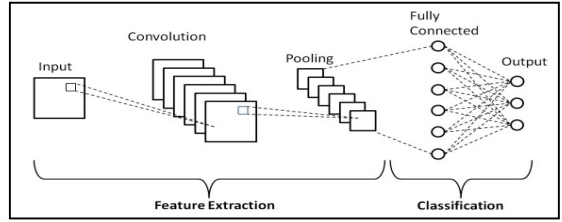

Fig. 2 Diagram of a Convolution Neural Network (Source: https://www.researchgate.net/figure/Schematic-diagram-of-a-basic-convolutional-neural-network-CNN-architecture-26_fig1_336805909)

Convolutional Neural Networks (CNN) are different from traditional neural networks (Feedforward Neural Networks) and recurrent neural networks. While recurrent neural networks are commonly used for natural language processing and speech recognition, convolutional neural networks are more often utilized for classification and computer vision tasks. The convolutional layer is the core building block of a CNN, and it is where the majority of computation occurs

### B. *Collaborative Filtering*

A recommender system is a subclass of Information Filtering System, that provides suggestions for items that most appropriate for a particular user, based on the data collected from a multitude of users. Recommender systems are particularly useful as they can help a user choose properly when there is an overwhelming number of options. Both the users and the services provided have benefited from these kinds of systems. The quality and decision-making process has also improved through these kinds of systems.

Collaborative Filtering is a technique used by recommender systems for making automated predictions about the interests of a user by collecting preferences from many users. In the more general sense, collaborative filtering is the process of filtering for information or patterns using techniques involving collaboration among multiple agents, viewpoints, data sources, etc. Collaborative filtering can be applied to various kinds of data including sensing and monitoring data, financial data etc. The overwhelming amount of data necessities mechanisms for efficient information filtering. Collaborative filtering is one of the techniques used for solving this problem.

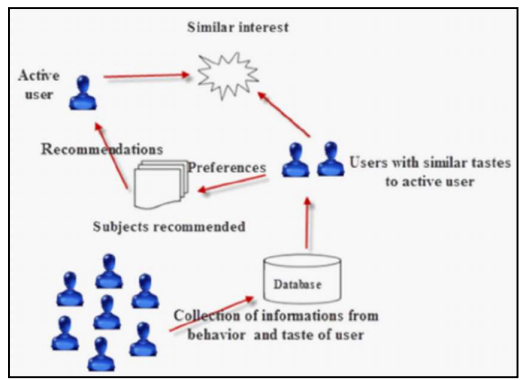

Fig. 3 Working of a Collaborative Filtering System (Source: https://www.researchgate.net/figure/Simple-description-work-of-collaborative-filtering_fig1_273268530)

Though various sources contradict, the discovery of the collaborative filtering algorithm is generally accredited to Dave Goldberg and his colleagues at Xerox PARC. The origins of modern recommender systems date back to the early 1990s when they were mainly applied experimentally to personal email and information filtering. Today, 30 years later, personalized recommendations are ubiquitous and research in this highly successful application area of AI is flourishing more than ever. Much of the research in the last decades was promoted by advances in machine learning technology. In 1992, the concept of "Collaborative Filtering" was introduced with an experimental mail system called Tapestry.

## II. THE CLASSICAL APPROACH

### A. Use of Logistic Classifiers

Collaborative Filtering is usually done using multivariate logistic regression as the output is discrete in nature (Logistic regression corresponds to discrete output, used in classification problems; while linear regression corresponds to continuous output, used in data-value prediction problems). With conventional regularized multivariate regression, appreciable accuracy is achieved. But, neural networks (perceptron) can help us to achieve considerably more accuracy using backpropagation and gradient descent. The classical approach to collaborative filtering using logistic regression is as follows:

$n_u = number\ of\ users$

$n_m = number\ of\ movies$

$r(i,j) = 1\ if\ 'j'\ has\ rated\ movie\ 'i'\ (0\ otherwise)$

$y^{(i,j)} = rating\ by\ user\ 'j'\ on\ movie\ 'i'\ (if\ defined)$

$\theta^{(j)} = parameter\ vector\ for\ user\ 'j'$

$x^{(i)} = feature\ vector\ for\ movie\ 'i'$

Given $x^{(1)}, \ldots, x^{(n_m)}$, estimate $\theta^{(1)}, \ldots, \theta^{(n_u)}$:

$$\min_{\theta^{(1)},\ldots,\theta^{(n_u)}} \frac{1}{2} \sum_{j=1}^{n_u} \sum_{i:r(i,j)=1} \left( \left(\theta^{(j)}\right)^T x^{(i)} - y^{(i,j)} \right)^2$$

Adding the regularization term to overcome overfitting or underfitting (due to high bias or variance):

$$\min_{\theta^{(1)},\ldots,\theta^{(n_u)}} \frac{1}{2} \sum_{j=1}^{n_u} \sum_{i:r(i,j)=1} \left( \left(\theta^{(j)}\right)^T x^{(i)} - y^{(i,j)} \right)^2 + \frac{\lambda}{2} \sum_{j=1}^{n_u} \sum_{k=1}^{n} \left(\theta_k^{(j)}\right)^2$$

Where $\theta \in \mathbb{R}^n$ i.e., $\theta$ is an $n-dimensional\ vector$, $\lambda = regularization\ parameter$
Similarly, Given $\theta^{(1)}, \ldots, \theta^{(n_u)}$ estimate $x^{(1)}, \ldots, x^{(n_m)}$:

$$\min_{x^{(1)},\ldots,x^{(n_m)}} \frac{1}{2} \sum_{i=1}^{n_m} \sum_{i:r(i,j)=1} \left( \left(\theta^{(j)}\right)^T x^{(i)} - y^{(i,j)} \right)^2 + \frac{\lambda}{2} \sum_{i=1}^{n_m} \sum_{k=1}^{n} \left(x_k^{(i)}\right)^2$$

Where $x \in \mathbb{R}^n$ i.e., $x$ is an $n-dimensional\ vector$

Considering the above functions as the cost functions for the respective parameter vectors ($\theta$) and feature vectors ($x$) as $J(\theta^{(1)}, \ldots, \theta^{(n_u)})$ and $J(x^{(1)}, \ldots, x^{(n_m)})$ we simultaneously minimize the functions using gradient descent. Thus, for every $j = 1,2,\ldots,n_u$ and $i = 1,2,\ldots,n_m$:

$$x_k^{(i)} = x_k^{(i)} - \alpha \left( \sum_{j:r(i,j)=1} \left( \left(\theta^{(j)}\right)^T x^{(i)} - y^{(i,j)} \right) \theta_k^{(j)} + \frac{\lambda}{2} x_k^{(i)} \right)$$

$$\theta_k^{(j)} = \theta_k^{(j)} - \alpha \left( \sum_{i:r(i,j)=1} \left( \left(\theta^{(j)}\right)^T x^{(i)} - y^{(i,j)} \right) x_k^{(i)} + \frac{\lambda}{2} \theta_k^{(j)} \right)$$

Where $x_k^{(i)}$ is the k[th] feature of the feature vector $x^{(i)}$ and $\theta_k^{(j)}$ is the k[th] parameter of the parameter vector $\theta^{(j)}$. The learning rate for the gradient descent is $\alpha$ and the regularization parameter for the same is $\lambda$. Both are hyperparameters and can be optimized using hyperparameter tuning.

After minimizing the cost functions and coming up with a set of feature vector ($X$) and parameter vector ($\theta$), we compute the hypothesis $h(x)$ using the sigmoid function

$$h(X) = \frac{1}{1+e^{-\theta^T X}}$$

where $X$ is the feature vector of the movie to be recommended to the user and $\theta$ is the parameter vector of the user. $h(x)$ is the estimated likelihood that user $\theta$ will like movie $X$

### B. A Better Alternative

Neural networks are complex machine learning models designed to fit complex datasets using backpropagation and gradient descent (or other advanced optimization algorithms). This can help us to achieve higher levels of accuracy in predicting the choice/taste of a user. Coupled with optimization techniques such as cosine similarity, gradient boosting and dimensionality reduction, it will be able to provide a more accurate outcome.

As seen in a number of experiments and research studies, artificial neural networks tend to provide a better overall fit to the data, thanks to its ability to form more complex decision boundaries. Thus, in this use case too, it would help us to find a better fit for the data.

## III. USING ARTIFICIAL NEURAL NETWORKS

### A. Objective of the Model

Given $x^{(1)}, \dots, x^{(n_m)}$ and $y^{(i,j)}$, estimate $\theta^{(1)}, \dots, \theta^{(n_u)}$
Compute: $Argmin\ C(\theta);\ C(\theta) \rightarrow$ Cost/Loss Function
Preprocessing of the dataset:

$$y^{(i,j)} = \begin{cases} 1: y^{(i,j)} \geq T \\ 0: y^{(i,j)} < T \end{cases} ; T \rightarrow Threshold\ of\ recommendation$$

The value of T can be set based on the rating of the movies. It signifies that the value of y will be set to 1, if the rating of the movie according to a user is greater than the threshold and 0, if the rating is lower than the same.

$$Cost(h(x), y) = \begin{cases} -\log h(x)\ ; y = 1 \\ -\log(1-h(x)); y = 0 \end{cases}$$

$$where,\quad h(x) = \frac{1}{1+e^{-\theta^T X}}$$

$\theta^T$ = Transpose of the matrix containing the parameter vectors of the user
X = Matrix containing the feature vectors of the movie
Combined form of the cost function:

$$cost(h(x), y) = -y \log(h(x)) - (1-y) \log(1 - h(x))$$

For artificial neural networks the total cost, denoted by $C(\theta)$ is:

$$C(\theta) = -\frac{1}{n_m} \left( \sum_{i=1}^{n_m} y^{(i)} \log\left(h(x^{(i)})\right) + (1 - y^{(i)}) \log\left(1 - h(x^{(i)})\right) \right)$$

For a neural network containing K output nodes, the cost/loss function gets modified. The sum over all the output nodes is taken in this case, and the function is modified in the following way:

$K \rightarrow number\ of\ ouput\ nodes\ of\ the\ neural\ network$

$$C(\theta) = -\frac{1}{n_m} \left( \sum_{i=1}^{n_m} \sum_{k=1}^{K} y_k^{(i)} \log\left(h(x^{(i)})_k\right) + (1 - y_k^{(i)}) \log\left(1 - h(x^{(i)})_k\right) \right)$$

Therefore, objective: compute Argmin $C(\theta)$:

$$Compute:\quad Argmin -\frac{1}{n_m} \left( \sum_{i=1}^{n_m} \sum_{k=1}^{K} y_k^{(i)} \log\left(h(x^{(i)})_k\right) + (1 - y_k^{(i)}) \log\left(1 - h(x^{(i)})_k\right) \right)$$

This will return the parameters $\theta^{(1)}, \ldots, \theta^{(n_u)}$ of a particular user which can be used with the feature vector of a new movie (one which user has not seen) to compute $h(x)$ and determine the likelihood of recommendation. If $h(x) \geq T$; T being the threshold of recommendation, then the movie can be recommended to the user. On the other hand, if $h(x) < T$, then the movie may not be a right fit for the particular user taken into consideration.

### B. *Working of the Proposed Model*

$g(x) = Activation\ function\ of\ the\ neural\ network$

The activation function helps to map the resulting values in between 0 to 1 or -1 to 1 etc. (depending on the function). Some of the activation functions that can be used in the neural network including linear and non-linear activation function are:

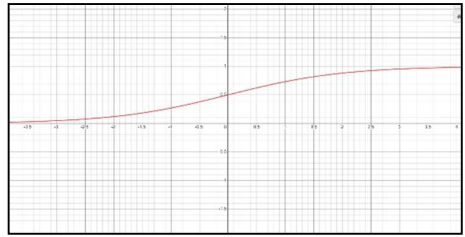

Fig. 4 Linear Graph representing the Sigmoid Function

Sigmoid/logistic activation function: $g(x) = \frac{1}{1+e^{-x}}$
Range: 0 to 1
Use Case: Since probability of an event lies between 0 and 1, the sigmoid function is especially useful for models where the prediction of a probability has to me made.

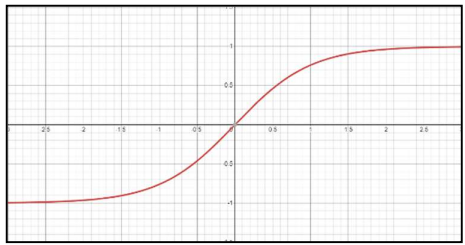

Fig. 5 Linear Graph representing the Hyperbolic tangent activation function

Hyperbolic tangent activation function: $g(x) = \tanh(x)$
Range: -1 to 1
Use Case: Mainly used for classification between two classes

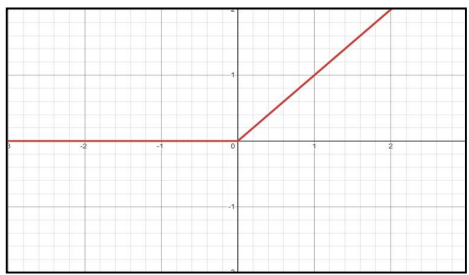

Fig. 6 Linear Graph representing the Rectifier Linear Unit (ReLU) activation function

Rectifier Linear Unit (ReLU) activation function: $g(x) = \max(0, x)$
Range: 0 to + infinity
Use Case: Most used activation function nowadays in neural networks.

Some used notation:
$a_i^{(j)}$ = activation of $i^{th}$ node of layer $j$
$U^z = \{\theta^{(n)}: 1 \leq n \leq j; j = $ number of layers of neural network$\}$; $U^z \to$ set of all matrices of weights for user $z$
$\theta^{(j)}$ = matrix of weights controlling function mapping from layer $j$ to $j+1$

$$\theta^{(j)} = \begin{bmatrix} \theta_{11} & \theta_{12} & \cdots & \theta_{1n} \\ \theta_{21} & \theta_{22} & \cdots & \theta_{2n} \\ \vdots & \vdots & \ddots & \vdots \\ \theta_{n1} & \theta_{n2} & \cdots & \theta_{nn} \end{bmatrix} = Matrix\ of\ user\ parameters$$

Considering the following neural network for our demonstration purposes:

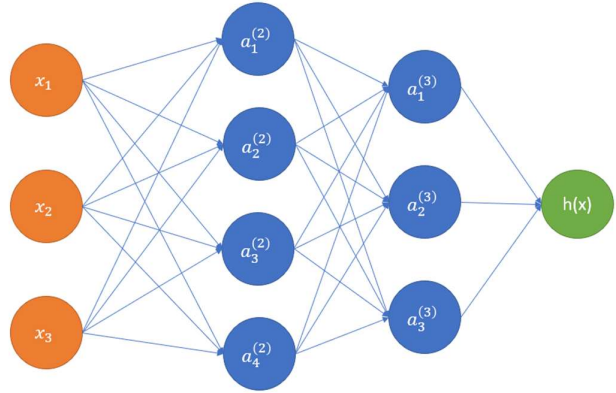

Where $X^{(1)} = \begin{bmatrix} x_1 \\ x_2 \\ x_3 \end{bmatrix}$ is the feature vector of the movie and first layer of the neural network and $a^{(2)} = \begin{bmatrix} a_1^{(2)} \\ a_2^{(2)} \\ a_3^{(2)} \\ a_4^{(2)} \end{bmatrix}$ is the second layer of the neural network. $a^{(2)}$ is also the first activation layer of the neural network. Similarly, the third layer/ second activation layer
of the neural network is $a^{(3)} = \begin{bmatrix} a_1^{(3)} \\ a_2^{(3)} \\ a_3^{(3)} \end{bmatrix}$

Using forward propagation, we calculate the values of the activation nodes of the first activation layer:

$$a_1^{(2)} = g(\theta_{11}^{(1)}x_1 + \theta_{12}^{(1)}x_2 + \theta_{13}^{(1)}x_3)$$
$$a_2^{(2)} = g(\theta_{21}^{(1)}x_1 + \theta_{22}^{(1)}x_2 + \theta_{23}^{(1)}x_3)$$
$$a_3^{(2)} = g(\theta_{31}^{(1)}x_1 + \theta_{32}^{(1)}x_2 + \theta_{33}^{(1)}x_3)$$
$$a_4^{(2)} = g(\theta_{41}^{(1)}x_1 + \theta_{42}^{(1)}x_2 + \theta_{43}^{(1)}x_3)$$

Therefore, the general formula for $i^{th}$ activation node of the second layer/ first activation layer ($j = 2$) of the neural network
having 'n' input nodes i.e., an n-dimensional vector containing 'n' features of the movie is:

$$a_i^{(2)} = g\left(\sum_{k=1}^{n} \theta_{ik}^{(1)} x_k\right); j = 2$$

Similarly for $1^{st}$ activation node of the third layer/ $2^{nd}$ activation layer of the neural network is:

$$a_1^{(3)} = g(\theta_{11}^{(2)}a_1^{(2)} + \theta_{12}^{(2)}a_2^{(2)} + \theta_{13}^{(2)}a_3^{(2)} + \theta_{14}^{(2)}a_4^{(2)})$$

Therefore, for the $k^{th}$ activation node of the layer 'j' having 'n' activation nodes in the layer 'j-1', for $j \geq 3$, the general formula
for the calculation of activation is as follows:

$$a_k^{(j)} = g\left(\sum_{i=1}^{n} \theta_{ki}^{(j-1)} a_i^{(j-1)}\right); j \geq 3$$

Therefore, for a neural network containing 'j' layers and having 'n' nodes in the layer 'j', the final hypothesis is given by:

$$h(x) = g\left(\theta_{11}^{(j)} a_1^{(j)} + \theta_{12}^{(j)} a_2^{(j)} + \theta_{13}^{(j)} a_3^{(j)} + \cdots + \theta_{1n}^{(j)} a_n^{(j)}\right) = g\left(\sum_{k=1}^{n} \theta_{1k}^{(j)} a_k^{(j)}\right)$$

If we consider the vectorized implementation of the above approach, then $a^{(2)} = g(\theta^{(1)} X^{(1)})$. Similarly, $a^{(3)} = g(\theta^{(2)} a^{(2)})$ and so on. Thus, the general vectorized implementation of activation for layer 'j' is:

$$a^{(j)} = g(\theta^{(j-1)} a^{(j-1)}); j \geq 3$$

Gradient computation using Backpropagation algorithm:
Total number of layers = L
$y^{(i,j)}$ = rating by user 'j' on movie 'i' (if defined)

For all $l \in L$ and $i, j \in \theta^{(n)} \forall \theta^{(n)} \in U^Z$, deviation factor, $\Delta_{ij}^l = 0$

Now, $\delta^{(L+1)} = h(x) - y^{(i,j)}$

Converting $\delta^{(L+1)}$ to a matrix to back-feed the neural network $\Rightarrow \delta^{(L+1)} = \begin{bmatrix} h(x) - y^{(i,j)} \\ \vdots \\ n \text{ times} \end{bmatrix}$; $n \to$ number of nodes for layer L

Note: Only required for layer L+1 i.e., the output node as the output node, h(x) outputs a definite value. So it needs to be converted into a matrix in order to carry on with the backpropagation gradient computation.

For all layers, $l \in \{L, L-1, \ldots 2\}$:

$$\delta^{(l)} = \left((\theta^{(l)})^T \delta^{(l+1)}\right) .* a^{(l)} .* (1 - a^{(l)})$$

Where ".*" operation means the element wise multiplication of the corresponding matrices.
Now, calculating the deviation factors,

$$\Delta_{ij}^{(l)} = \Delta_{ij}^{(l)} + a_j^{(l)} \delta_i^{(l+1)}$$

A more general vectorized implementation is $\Delta^{(l)} = \Delta^{(l)} + \delta^{(l+1)} (a^{(l)})^T$

$$D_{ij}^{(l)} = \frac{1}{m} \Delta_{ij}^{(l)}; where \: D \to partial \: derivative \: of \: the \: term$$

For each layer, using matrices, $D^{(l)} = \frac{1}{m} \Delta^{(l)}$; m = total number of movie feature vectors
Through calculus, it can be shown,

$$\frac{\partial}{\partial \theta_{ij}^{(l)}} C(\theta) = D_{ij}^{(l)}; \: C(\theta) \to Cost \: function \: of \: the \: neural \: network$$

Therefore, $D^{(l)}$ is the vector containing the partial derivatives of the activation nodes for the layer $l$

$$Gradient \: matrix, G = \begin{bmatrix} i \: \forall \: i \in D^{(1)} \\ i \: \forall \: i \in D^{(2)} \\ \vdots \\ i \: \forall \: i \in D^{(L)} \end{bmatrix}$$

Thus, G is the gradient matrix containing all the weights for the neural network.
Now running gradient descent on the cost function $C(\theta)$ and minimizing the weights in the gradient matrix, we obtain the matrix containing the parameter vectors of the user.
Therefore, we compute:

$$Argmin - \frac{1}{n_m} \left(\sum_{i=1}^{n_m} \sum_{k=1}^{K} y_k^{(i)} \log\left(h(x^{(i)})_k\right) + (1 - y_k^{(i)}) \log\left(1 - h(x^{(i)})_k\right)\right)$$

Where $h(x)$ is computed using the weights present in the gradient matrix.

The model thus develops the parameter vectors of the user and can predict whether the user will like a particular movie which is not in the training dataset using the feature vectors of the movie.

## C. Optimization Techniques

The following optimization techniques can be implemented in the model to increase the computational efficiency of the same as well as to achieve a better overall result.

1) **Implementing Gradient Checking**: Backpropagation is very powerful if implemented correctly. While building a neural network from scratch, backpropagation is often the place where people make mistakes. Implementing backpropagation incorrectly may not only result in the improper estimation of the weights, but also the total failure of the neural network. Thus, an intermediate step, known as gradient checking could help in overcoming this problem. This is a powerful way to eliminate all the bugs in backpropagation.

   From calculus we know,
   $$f'(x) = \frac{d}{dx}f(x) = \lim_{h \to 0} \frac{f(x+h) - f(x-h)}{2h}$$

   Using this, we can conduct the numerical estimation of the gradient by
   $$\frac{d}{d\theta}C(\theta) \approx \frac{C(\theta + \gamma) - C(\theta - \gamma)}{2\gamma}; where\ \gamma \approx 10^{-7}$$

   Therefore, $\forall\ \theta_i \in \{\theta\ \forall\ \theta \in \theta^{(l)}\}$:
   $$\frac{\partial}{\partial \theta_i}C(\theta) \approx \frac{J(\theta_1, \theta_2, \dots, \theta_i + \gamma, \theta_{i+1}, \dots \theta_n) - J(\theta_1, \theta_2, \dots, \theta_i - \gamma, \theta_{i+1}, \dots \theta_n)}{2\gamma}$$

   Let the matrix formed by the numerical estimation of the gradients be $G'$. Now we calculate the Euclidian distance normalized by the sum of the sum of the vectors $G$ and $G'$.
   $$\varepsilon = \frac{\|G' - G\|_2}{\|G'\|_2 + \|G\|_2}$$

   If $\varepsilon < \gamma$, we can conclude that the backpropagation was implemented correctly and is therefore working.
   Note: While gradient checking is an incredibly powerful way of checking whether backpropagation is working properly, it must be noted that it is a very memory intensive process and therefore must be implemented only once i.e., after the initial calculation of the gradient matrix using backpropagation. If gradient checking is allowed to run after every iteration of gradient descent, then the efficiency of the model will be hampered.

2) **Random Initialization of the Parameters:** While setting all the initial value of the weights to zero works for logistic regression, it doesn't work for neural networks. This is because, for all iterations, the weights of the activation nodes will be equal. This will therefore result in the parameters going into each of the nodes be equal. That is,
   $$\frac{\partial}{\partial \theta^1_{11}}C(\theta) = \frac{\partial}{\partial \theta^1_{12}}C(\theta) = \dots = \frac{\partial}{\partial \theta^1_{1n}}C(\theta)$$

   Thus, the neural network always ends up with only one feature. This will result in high bias and will therefore result in under-fitting of the parameters.

   Symmetry breaking: Initialize every $\theta \in \theta^{(l)}\ \forall\ \theta^{(l)} \in U^Z$ to some random value in between $[-x.x]$ e.g., 0.69420

   Note: Usually it is a good practice to randomize all the weights in the range (0, 1), such that every weight is a random real number between 0 and 1 (both exclusive). The weights can also be between -1 and 1 (both exclusive).

3) **Feature Scaling:** The idea is to make sure that all the features are on a similar scale. Gradient descent can converge faster if the features are in a similar range. Therefore, for all the different features, the features need to be divided by some constant (maybe be different for different features) to get them into approximately -1⩽x⩽1 range.

4) **Mean Normalization:** Just like feature scaling, this falls under the category of preprocessing of data. Here the idea is to make all the features have zero mean. Having a normalized dataset helps the gradient descent to converge faster, thereby lowering the training time of the model.

$$x_i = \frac{x_i - \mu_i}{\max x_i - \min x_i}; \mu_i \rightarrow mean\ of\ the\ dataset$$

Replacing the denominator by standard deviation instead, we obtain a similar result. The process is called mean standardization.

5) **Implementing Regularization:** A machine learning model is often prone to high bias or high variance. The former happens when the hypothesis functions maps poorly to the trend of the data and is also known as under-fitting. The latter is when the learned hypothesis may fit the training set very well but fail to generalize to new examples outside the training dataset. This can be eliminated by keeping all the features but reducing the magnitude/values of the parameters.

Therefore, we penalize the cost function using a new introduced term, λ, known as regularizing parameter, thereby minimizing the values of the parameters.

The regularized cost function for the neural network is:

$$C(\theta) = -\frac{1}{n_m}\left(\sum_{i=1}^{n_m}\sum_{k=1}^{K} y_k^{(i)} \log\left(h(x^{(i)})_k\right) + (1-y_k^{(i)}) \log\left(1-h(x^{(i)})_k\right)\right) + \frac{\lambda}{2n_m}\sum_{l=1}^{L-1}\sum_{i=1}^{S_l}\sum_{j=1}^{S_{j+1}}\left(\theta_{ji}^{(l)}\right)^2$$

Where $S_l, S_{l+1}$ = number of units in layer $l, l+1$ respectively; $L$ = total number of layers in neural network

Note: If the value of λ i.e., the regularization parameter is too large, all the parameters are close to zero. This results in underfitting i.e., the hypothesis has too high bias. Similarly, a too small value of the regularizing parameter will make the regularization term very small and thus will result in regularization becoming useless. Thus, we should pick a moderate value of the regularization parameter (around 10) and modify it overtime to find the best fit.

6) **Hyperparameter Tuning**: In machine learning, a hyperparameter is a parameter whose value is used to control the learning process unlike the other parameters whose values are derived via training. For the neural network, we have two hyperparameters involved i.e., the learning rate, α, of the gradient descent and the regularization parameter, λ, for the regularized cost function.

There are various ways of going about hyperparameter tuning. The most common ones are GridSearchCV and RandomizedSearchCV. In GridSearchCV approach, the machine learning model is evaluated for a range of hyperparameter values. It is called GridSearchCV because it searches for the best set of hyperparameters from a grid of hyperparameter values.

RandomizedSearchCV solves a drawback of GridSearchCV, as it goes through only a fixed number of hyperparameter settings. It moves within the grid in a random fashion to find the best set of hyperparameters. This approach reduces unnecessary computation.

Note: For hyperparameter tuning, the RandomSearchCV and GridSearchCV functions can be used from sklearn.model_selection module

7) **Using Stochastic Gradient Descent or Mini-Batch Gradient Descent:** Stochastic gradient descent is selecting data points at each step to calculate the derivatives. Stochastic gradient descent randomly picks one data point from the whole dataset at each iteration to reduce the computations enormously.

The way of going about stochastic gradient descent is as follows:

$$cost\left(\theta,\left(x^{(i)}, y^{(i)}\right)\right) = \frac{1}{2}\left(h(x^{(i)}) - y^{(i)}\right)^2$$

Therefore, calculating the derivative of the cost function, we have the following expression:

$$\frac{\partial}{\partial \theta_j} cost\left(\theta, \left(x^{(i)}, y^{(i)}\right)\right) = \left(h(x^{(i)}) - y^{(i)}\right)x_j^{(i)}$$

Thus, the following steps should be taken in order to implement Stochastic Gradient Descent (SGD):

1. Randomly shuffle the dataset. Randomly shuffled dataset at first will help the gradient descent converge a little faster to the global minimum.

2. Repeat {

    for i = 1,2…, m {

    $\theta_j = \theta_j - \alpha\left(\left(h(x^{(i)}) - y^{(i)}\right)x_j^{(i)}\right); for\ all\ j = 0, 1, …, n\ (n = total\ number\ of\ features)$

    }

}

Looking through each example and take a little step to try to fit just that particular parameter.

For mini-batch gradient descent we define a parameter "b", the mini batch size i.e., using "b" examples for every iteration. Usually the value of "b" is taken somewhere between 2 and 100. The process is as follows:

Repeat {

Get "b" examples → $\left(x^{(i)}, y^{(i)}\right), …, \left(x^{(i+(b-1))}, y^{(i+(b-1))}\right)$

$\theta_j = \theta_j - \alpha\frac{1}{b}\sum_{k=i}^{i+(b-1)}\left(h(x^{(k)}) - y^{(k)}\right)x_j^{(k)}; for\ all\ j = 0,1, …, n\ (n = total\ number\ of\ features)$

i = i + b

}

Using mini-batch gradient descent results in more stable convergence towards the global minimum since we are calculating an average gradient over "b" examples.

Note: While mini-batch gradient descent is computationally more efficient, the one major turnoff in this case is the introduction of a new hyperparameter "b", i.e., the batch size. Thus, to ensure peak efficiency, we have to tune the mini batch size, "b", as well along with the other hyperparameters using hyperparameter tuning.

The above optimization techniques help us not only to eliminate bugs and improve accuracy of the model, but also goes a long way in improving the computational efficiency of the same. Therefore, we should implement all of the above techniques to help us make the model optimal.

## IV. CONCLUSION

While logistic regression classifiers provide a fairly accurate result of the user's recommendation based on the data collected from a multitude of users, we can further enhance the accuracy of the model by the introduction of neural networks. As neural networks are designed to fit complex datasets, we can form functions of higher complexity and thus get a better fit for the data without running into issues like high bias (under-fitting) and high variance (over-fitting). While neural networks can fit the datasets better than logistic classifiers, they are often computationally more expensive. Thus, we implement a series of optimization techniques to get the best possible results with the least computational work. This includes preprocessing of data i.e., implementing mean normalization and feature scaling. While training the model, we use gradient checking to ensure backpropagation is implemented correctly, we use regularization to overcome the issue of under-fitting and over-fitting due to high bias and high variance respectively. We initialize the parameter vectors using random initialization. To make the model computationally more efficient, instead of regular batch gradient descent, we use stochastic/mini-batch gradient descent. Finally, to achieve the best set of hyperparameters i.e., the learning rate ($\alpha$), the regularization parameter ($\lambda$) and the mini-batch size (b), we use hyperparameter tuning using GridSearchCV or RandomizedSearchCV. More advanced algorithms such as Momentum (used for reducing the high variance in SGD and softening the convergence), Nesterov Accelerated Gradient (made so that

Momentum does not miss the local minima), AdaGrad (overcomes the problem of having a constant learning rate and therefore implements a dynamic learning rate), AdaDelta (removes the decaying learning rate problem of AdaGrad) and Adaptive Momentum Estimation (works with momentums of first and second order) can be used instead of gradient descent to further improve the computational efficiency and accuracy. For neural networks specifically, Adaptive Momentum Estimation tends to be the best optimizer.

The objective (section III-A) and the working (section III-B) of the model have been explained with the specific example of recommending a movie to a user. However, the similar approach can be followed in other use cases as the core concept remains the same. The most important part is the implementation of gradient checking just after the calculation of the derivatives using backpropagation to ensure the proper working of the backpropagation algorithm. The perceptron collaborative filtering can be used in OTT platforms to recommend movies to a user based on the rating of other users. It can also be used in online shopping platforms to recommend certain products to certain people based on relatability and rating/reviews.

ACKNOWLEDGEMENT

First, I would like to thank my parents and family members for their constant support. Next, I would like to thank all the teachers of the computer science department of Delhi Public School, Ruby Park for understanding my potential and helping to increase the same. Finally, I am heavily indebted to all the editors of the journal for their time and effort in going through the research paper.

REFERENCES

[1]  Ulrika Jägare, "*Data Science Strategy*"
[2]  U. Dinesh Kumar, "*The Science of Data-Driven Decision Making*"
[3]  Manaranjan Pradhan and U. Dinesh Kumar, "*Machine Learning using Python*"
[4]  Abraham Silberschatz, Peter B. Galvin and Greg Gagne, "*Operating System Concepts*"
[5]  Richard P. Feynman, "*Feynman Lectures on Computation*"
[6]  What is a Neural Network, https://www.investopedia.com/terms/n/neuralnetwork.asp, last accessed on 2023-02-03
[7]  Hopfield Network, https://en.wikipedia.org/wiki/Hopfield_network, last accessed on 2023-02-03
[8]  Hopfield Neural Network, https://www.geeksforgeeks.org/hopfield-neural-network/, last accessed on 2023-02-03
[9]  Hopfield Network, https://www.javatpoint.com/artificial-neural-network-hopfield-network, last accessed on 2023-02-03
[10] What are recurrent neural networks, https://www.ibm.com/topics/recurrent-neural-networks, last accessed on 2023-02-03
[11] Backpropagation, https://en.wikipedia.org/wiki/Backpropagation, last accessed on 2023-02-03
[12] History of Artificial Neural Network, https://www.javatpoint.com/history-of-artificial-neural-network, last accessed on 2023-02-03
[13] Convolutional Neural Networks, https://www.ibm.com/topics/convolutional-neural-networks, last accessed on 2023-02-03
[14] What Are Recommendation Systems in Machine Learning, https://www.analyticssteps.com/blogs/what-are-recommendation-systems-machine-learning, last accessed on 2023-02-03
[15] Recommender system, https://en.wikipedia.org/wiki/Recommender_system, last accessed on 2023-02-03
[16] United we find, https://www.economist.com/technology-quarterly/2005/03/12/united-we-find, last accessed on 2023-02-03
[17] *Recommender Systems: Past, Present and Future*, https://ojs.aaai.org/index.php/aimagazine/article/view/18139, last accessed on 2023-02-03, DOI: 10.1609/aaai.12012
[18] Activation Functions in Neural Networks, https://towardsdatascience.com/activation-functions-neural-networks-1cbd9f8d91d6, last accessed on 2023-02-03
[19] How to Debug a Neural Network with Gradient Checking, https://towardsdatascience.com/how-to-debug-a-neural-network-with-gradient-checking-41deec0357a9, last accessed on 2023-02-03
[20] The Distance Between Two Vectors, http://mathonline.wikidot.com/the-distance-between-two-vectors, last accessed on 2023-02-03
[21] Regularization Parameter, https://www.sciencedirect.com/topics/engineering/regularization-parameter, last accessed on 2023-02-03
[22] Hyperparameter (machine learning), https://en.wikipedia.org/wiki/Hyperparameter_(machine_learning), last accessed on 2023-02-03
[23] Stochastic Gradient Descent, https://towardsdatascience.com/stochastic-gradient-descent-clearly-explained-53d239905d31, last accessed on 2023-02-03
[24] Various Optimization Algorithms for Training Neural Network, https://towardsdatascience.com/optimizers-for-training-neural-network-59450d71caf6, last accessed on 2023-02-03
[25] Andrew Ng, "*Machine Learning Yearning*"